\documentclass[preprint]{aastex}

\shorttitle{Coronal Rain Dynamics}
\shortauthors{Fang et al.}

\begin{document}

\title{Coronal rain in magnetic arcades: Rebound shocks, Limit cycles, and Shear flows}
\author{X.~Fang, C.~Xia, R.~Keppens and T.~Van Doorsselaere}
\affil{Centre for mathematical Plasma Astrophysics, Department of Mathematics, KU Leuven, Celestijnenlaan 200B, 3001 Leuven, Belgium}

\begin{abstract}
We extend our earlier multidimensional, magnetohydrodynamic simulations of coronal rain occurring in magnetic arcades with higher resolution, grid-adaptive computations covering a much longer ($>6$ hour) timespan.
We quantify how in-situ forming blob-like condensations grow along and across field lines and show that rain showers can occur in limit cycles, here demonstrated for the first time in 2.5D setups. We discuss dynamical, multi-dimensional aspects of the rebound shocks generated by the siphon inflows and quantify the thermodynamics of a prominence-corona-transition-region like structure surrounding the blobs. We point out the correlation between condensation rates and the cross-sectional size of loop systems where catastrophic cooling takes place. We also study the variations of the typical number density, kinetic energy and temperature while blobs descend, impact and sink into the transition region. In addition, we explain the mechanisms leading to concurrent upflows while the blobs descend. As a result, there are plenty of shear flows generated with relative velocity difference around 80 km s$^{-1}$ in our simulations. These shear flows are siphon flows set up by multiple blob dynamics and they in turn affect the deformation of the falling blobs. In particular, we show how shear flows can break apart blobs into smaller fragments, within minutes.
\end{abstract}

\keywords{magnetohydrodynamics(MHD) --- Sun: corona --- Sun: filaments, 
prominences}

\section{Introduction}
Observations show recurrent formation and reshuffling of cool and dense material in coronal loops. The small scale (${\cal O}(100)$ km) coronal rain is observed as cold, dense elongated blob-like features condensing in a much hotter loop and descending along one of its legs. The rain is guided by the loop magnetic field \citep{beckers62}, dropping from heights of tens of Mm into the chromosphere \citep{kawaguchi70,leroy72}. Similar phenomena have been observed by analysing absorption profiles in EUV spectral lines \citep{schrijver01,oshea07}. Seen to propagate from the top of the loop towards its footpoints \citep{groof04}, systematic intensity variations in EUV spectral lines are confirmed as downflows of cool plasma, rather than representing slow magneto-acoustic waves \citep{groof05}. Besides downflows towards footpoints, \citet{tripathi09} also found upflows towards the loop apex. \citet{antolin10} observed and tracked coronal rain in 30 active region loops and found more than one hundred descending condensations within 71 minutes. Tracing the cool material towards loop footpoints, \citet{kamio11} observed propagating patterns suggesting a hot upflow following the downflows, supplying hot plasma into the loops. \citet{antolin12} suggested that coronal rain consists of plenty of small blobs, with sizes around 300 km in width and 700 km in length on average and they also estimated the occurrence rate of coronal rain in active region loops to be once every two days. Since the solar corona is swamped with magnetic loops, this suggests a scenario where coronal rain is rather common.

Considering the very small sizes involved, one of the most attractive features of coronal rain is that it can be used to probe the local magnetic field structure, or that it can expose valuable properties of the local thermodynamic conditions inside coronal loops \citep{antolin10}. Indeed, the magnetic field structure has a much longer lifetime than the timescale for condensations to form and fall \citep{beckers62}. Additionally, due to the low temperature (of order $10^{4}$ K) of these condensations, coronal rain is normally observed in cold chromospheric lines \citep{levine77,muller05}. Coronal rain results from thermal instability, with its non-linear counterpart and evolution also known as thermal non-equilibrium or catastrophic cooling. The linear thermal instability takes places whenever radiative losses locally overcome the heating input and is governed by well-known stability criteria for uniform radiative plasma conditions \citep{parker53,field65}. These can be met in the coronal temperature range, as one encounters locally negative slopes in the radiative loss function $\Lambda$(T) as function of temperature. When thermal conduction is insufficient in transporting enough energy to cooling (and condensing) material, the temperature reduces over time. As a consequence of thermal instability, temperature and gas pressure drop dramatically in the perturbed region, resulting in matter sucked in from the surroundings to the perturbed region, forming an increasingly larger and cooler condensation. This runaway effect will continually increase the density and decrease the temperature of condensations until heating and cooling achieve a balance again at lower temperatures and higher densities.Numerical simulations have contributed to our understanding of thermal instability over the last 40 years \citep{goldsmith71,hildner74,mok90,antiochos91,dahlburg98,antiochos99,antiochos00,karpen03,karpen06,mok08,karpen08,xia11}. Early numerical work shows that in the million degrees solar corona, small temperature contrasts could be enhanced by line and recombination-driven radiative losses within several minutes \citep{goldsmith71}. Catastrophic cooling drives recombination of elements in the cool condensations, making them partially ionized and visible in cool chromospheric lines. \citet{hildner74} concluded that the rate of condensation is determined by hydrodynamical processes mainly. 

An important progress in modeling was obtained in \citet{antiochos91}, by using a spatially dependent heating increase that is localized nearer to the chromospheric footpoints than to the loop midpoint. With this localized heating at the footpoints, \citet{dahlburg98} pointed out that another key requirement to generate a stable, prominence-like condensation is a dipped geometry in the loop. With an adaptive grid code, \citet{antiochos99} showed, in a 1D model, that the complete growth of a condensation reached a quasi-steady state after $\approx$ 5000 s. In a similar 1D setup, \citet{xia11} calculated the linear instability criterion from numerical results and proved that the onset of coronal condensation indeed satisfies the linear isochoric instability criterion \citep{parker53}. In the solar corona, the fairly high densities required for the instability onset are thought to be obtained by evaporating material with heating located near the footpoints of coronal loops in the chromosphere or by direct mass injection into the corona \citep{wang99,chae01}, resulting e.g. from nano-heating events.  

Influenced by magnetohydrodynamic forces (gravity, Lorentz force and gas pressure gradients), condensations, once formed, either fall from the corona down to the chromosphere as coronal rain or they collect into larger structures and remain suspended in the corona over long time periods as prominences, supported by the magnetic field. Many numerical works addressed formation and dynamics of coronal rain, but adopted simplifying one-dimensional (1D) approximations reducing the problem to gas dynamic, thermodynamic evolutions along individual field lines \citep{antiochos91,schrijver01,karpen01,karpen05,muller03,muller04,groof05,karpen06,mendoza05,antolin10,xia11,luna12}. Since coronal rain occurs in many active region loops, the heating input is generally thought to be concentrated at the loop footpoints \citep{antiochos91,mendoza05}, which evaporates chromospheric plasma into the loops and increases the density. With a persistent heating, the anisotropic thermal conduction and optically thin radiation lead these coronal hot loops to reach thermally unstable regimes with a higher density in a timescale of hours \citep{xia11}. Then catastrophic cooling sets in locally, resulting in the fast formation of cool condensations, as demonstrated in 1D models \citep{karpen01,muller03,groof05,karpen05}. Numerical simulations by \citet{muller04} emphasized that a loss of equilibrium at the loop apex and the process of catastrophic cooling is caused by constant heating concentrated at the footpoints of the loop rather than a drastic decrease of the total loop heating which was used in earlier models. \citet{muller05} compared observations from an EIT shutterless campaign with simulations of coronal loops and confirmed that observed localized brightenings and fast flows are consistent with this model. An important conclusion from \citet{antolin10} was that the structure and dynamics of the coronal rain blobs are more sensitive to the pressure variations arising from catastrophic cooling than to gravity itself. This is in agreement with \citet{schrijver01}, who suggested that the internal pressure evolution of the loops, rather than gravity, determines the condensation speeds. Furthermore, \citet{antolin10} indicated that if a loop is predominantly heated by Alfv\'en waves, coronal rain is inhibited since they tend to heat the loop uniformly. Hence, coronal rain may not only point to the spatial distribution of the heating in coronal loops but also to the agent of the heating itself. They thus propose coronal rain as a marker for coronal heating mechanisms. \citet{xia11} pointed out that steady heating is not necessary to sustain the condensation. Once the condensation is formed, it keeps growing even after localized heating ceases. \citet{luna12} simulated a three-dimensional sheared double arcade with a large ensemble of 1D independent flux tubes and observed the formation of both prominence threads and coronal rain.

Recently, \citet{fang13} presented the first multidimensional, magnetohydrodynamic simulations which captured the initial formation and the long-term sustainment of the coronal rain phenomenon. 
There we found that coronal rain in arcades is always accompanied by fast counter-streaming siphon flows in neighbouring flux bundles and we statistically analysed 80 minutes of virtual coronal rain in terms of sizes, mass, and velocity patterns. Our 2.5D simulations showed how blobs deform into V-shaped patterns, and had blobs that levitate, evaporate in-situ, or fall into the transition region at speeds below free-fall. 
IRIS data recently revealed also many coronal rain impact events, with up to supersonic speeds above sunspots \citep{kleint14}. We therefore revisited our MHD setup from \citet{fang13}, at even further increased numerical resolution and for much longer time, going up to 6 hours in total. We now analyse blob formation and blob impact into the transition region in more detail, focusing on multi-dimensional aspects not probed by 1D setups. Furthermore, the High-resolution Coronal Imager (Hi-C) in July 2013 provided a much more detailed look at the fine structure and dynamics in the solar corona. With data from Hi-C, \citet{alexander13} reported that anti-parallel flows have been directly imaged along fundamental filament threads within the million degree corona. They measured relative flow velocities of similar magnitude as in our previous simulations, namely 70-80 km s$^{-1}$. Both observations and our simulations hence suggest that such counter-streaming flows are likely commonplace. We observed that siphon flows establish naturally in a raining arcade, with velocity differences on adjacent field lines up to 80 km s$^{-1}$. We thus also extended our simulations to further argue how our setup in a low field (order 12 G) magnetic arcade relates to the observed clumps of falling coronal rain \citep{antolin10} and to unresolved fine-scale structure in solar coronal loop-tops \citep{scullion14}.

The paper is then organized as follows: in \textsection2 we describe the numerical setup, in \textsection3.1 we describe the multidimensional aspects of the condensations, focusing on rebound shocks and their Prominence-Corona-Transion Region (PCTR) structure,  \textsection3.2 discusses the condensation rates and the long term coronal rain limit cycle obtained,  \textsection3.3 quantifies blob impact on the transition region and concurrent upflows, in \textsection3.4 we investigate the counter-streaming flows, and in \textsection3.5 we describe the shear flow effects. Conclusions are drawn in \textsection4. 

\section{Computational Aspects}\label{setup}

\subsection{Governing Equations and Initial Setup}
Our numerical setup follows our previous 2.5D thermodynamic MHD simulation from \citet{fang13}, which includes gravity, field-aligned heat conduction and radiative cooling and parametrized heating terms, 
on a rectangular plane with horizontal extension -40 Mm $\leq  x \leq 40$ Mm and vertical extension 0 $\leq  y \leq 50$ Mm.
The governing equations are as follows:
\begin{equation}
 \frac{\partial{\rho}}{\partial{t}}+\nabla\cdot\left(\rho \textbf{v} \right)= 0,
\end{equation}
\begin{equation}
 \frac{\partial{\left(\rho\textbf{v}\right)}}{\partial{t}}+\nabla\cdot\left(
\rho\textbf{vv}+p_{tot}\textbf{I}-\frac{\textbf{BB}}{\mu_{0}} \right)=\rho \textbf{g},
\end{equation}
\begin{equation}
 \frac{\partial{E}}{\partial{t}}+\nabla\cdot\left(E\textbf{v}+p_{tot}\textbf{v}-
\frac{\textbf{v}\cdot\textbf{B}}{\mu_{0}}\textbf{B}\right)=\rho\textbf{g}\cdot\textbf{v}+
\nabla\cdot\left( \vec{\kappa}\cdot\nabla T\right)-Q+H,
\end{equation}
\begin{equation}
 \frac{\partial{\textbf{B}}}{\partial{t}}+\nabla\cdot\left(\textbf{vB}-
\textbf{Bv}\right)=0,
\end{equation}
where $T$, $\rho, \textbf{B}, \textbf{v}$, and $\textbf{I}$ are respectively temperature, density, magnetic field, velocity, and unit tensor, with the ratio
of specific heats $\gamma$=5/3, and a total energy density as $E=p/\left(\gamma-1\right)+\rho v^{2}/2+
B^{2}/2\mu_{0}$ ; $p_{tot}\equiv p+B^{2}/2\mu_{0}$ is the total
pressure, consisting of magnetic pressure and thermal pressure $p$; \textbf{g}$=g_{0}R^{2}_{\odot}/\left(
R_{\odot}+y\right)^{2}$\textbf{$\hat{y}$} is the gravitational acceleration
 with the solar surface gravitational acceleration $g_{0}=-274\,\mathrm{m}/\mathrm{s}^2$ and the
 solar radius $R_{\odot}$; $H$ and $Q$ are respectively the heating
and radiative loss rates; $\vec{\kappa}$ is the thermal conductivity tensor. Assuming
a 10:1 abundance of hydrogen and helium of completely ionized plasma, we
obtain $\rho=1.4m_{\rm{p}}n_{\rm{H}}$, where $m_{\rm{p}}$ is the proton mass and $n_{\rm{H}}$ is
the number density of hydrogen. We use the ideal gas law $p=2.3n_{\rm{H}}k_{\rm{B}}T$,
where $k_{\rm{B}}$ is the Boltzmann constant. We also adopt $Q=1.2 n^{2}_{\rm{H}}\Lambda
\left(T\right)$ as the radiative cooling term, where $\Lambda\left(T\right)$
is the radiative loss function for optically thin emission, quantified by \citet{colgan08} using a recommended set
of quiet-region element abundances, as used in our previous work \citep{xia11,xia12,fang13,keppens14a}. In the calculations, \citet{colgan08} used a complete and self-consistent atomic data set and an
accurate atomic collisional rate over a wide temperature range. 
Below 10,000 K, we set $\Lambda\left(T\right)$ to vanish because the plasma there is optically thick and no longer fully ionised. We use the exact integration method as introduced by \citet{townsend09} to evaluate the radiative loss term. The use of explicit, (semi-)implicit, and exact integration
methods in grid-adaptive simulations has been compared in \citet{marle11}.
The term containing $\vec{\mathbf{\kappa}}=\kappa_{||}\mathbf{\hat{b}}\mathbf{\hat{b}}$ quantifies the anisotropic
thermal conduction along the magnetic field lines, composed by the unit vector
$\mathbf{\hat{b}}$ along the magnetic field and the Spitzer conductivity $\kappa_{||}$
as $10^{-6} \textit{T}$ $^{5/2}$ erg cm$^{-1}$ s$^{-1}$ K$^{-3.5}$.

We employ a linear force-free magnetic field for the initial magnetic configuration, which is characterised by a constant angle $\theta_0$ as 
follows:
\begin{displaymath}
 B_{x}=-B_{0} \cos \left( \frac{\pi x}{L_{0}} \right) \sin\theta_0 \exp\left(
 -\frac{\pi y \sin\theta_0}{L_{0}} \right)\,,
\end{displaymath}
\begin{displaymath}
 B_{y}=B_{0} \sin \left( \frac{\pi x}{L_{0}} \right) \exp\left(
 -\frac{\pi y \sin\theta_0}{L_{0}} \right)\,,
\end{displaymath}
\begin{equation}
 B_{z}=-B_{0} \cos \left( \frac{\pi x}{L_{0}} \right) \cos\theta_0 \exp\left(
 -\frac{\pi y \sin\theta_0}{L_{0}} \right)\,.
\label{bfield}
\end{equation}
with $\theta_0=30^\circ$, the arcade makes a $30^\circ$ angle 
with the neutral line ($x = 0, y=0$). $L_{0}=80$ Mm is the horizontal size of our domain from -40 Mm to 40 Mm, and 
when adopting $B_{0}=12$ G, our magnetic arcade has a total box averaged field strength of 2.9 G.

\begin{figure}
 \centering
 \includegraphics[width=\textwidth]{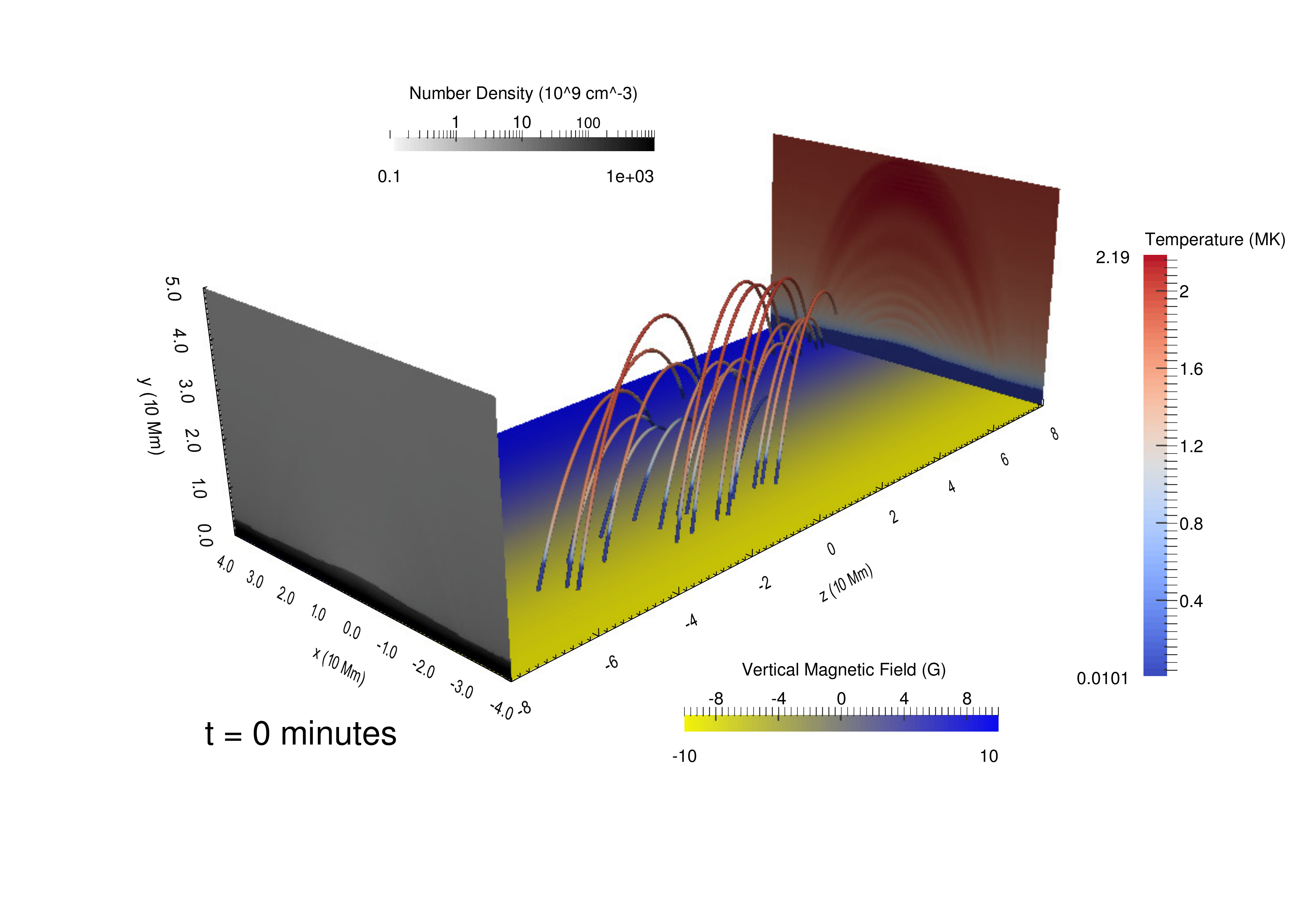}
 \caption{Around $t\approx 70$ minutes after relaxation, we show a 3D view on the quasi-equilibrium stage of our simulation, which serves as initial condition when extra localized heating is turned on. (the $t=0$ minutes means the resetting of time to zero from now on.) The randomly selected field lines are colored by temperature, the back cross-section shows the temperature while the front $x-y$ cross-section shows the number density map.}
 \label{3d}
\end{figure}

For the initial thermal structure, we set a uniform temperature of 10000 K below a height of 2.7 Mm and choose
a temperature profile with height ensuring a constant vertical
thermal conduction flux (i.e., $\kappa \partial{T}/\partial{y}$ = 2 $\times$ 10$^{5}$ erg cm$^{-2}$ s$^{-1}$)
above this height, as also exploited by other authors \citep{fontenla91,mok05}. The
initial density is then derived by assuming hydrostatic
equilibrium with the number density of 1.2 $\times$ 10$^{15}$ cm$^{-3}$ at the
bottom and the initial velocity field of all plasma is static. Since the corona needs to achieve a self-consistent thermal structure, we employ
a background heating rate decaying exponentially with height into the whole system all the time,
\begin{equation}
 H_{0}=c_{0} \exp\left(-\frac{y }{\lambda_{0}} \right) 
 \,,
\label{heat0}
\end{equation}
where $c_{0}=10^{-4}$ erg cm$^{-3}$ s$^{-1}$ and $\lambda_{0}=50$ Mm. This heating is meant to balance the 
radiative losses and heat conduction related losses of the corona in its steady state. 
The slight difference in heating scale height between 50 Mm in equation~(\ref{heat0}) above and 36 Mm in equation (2) in \citet{fang13}  
improves numerical stability at the top boundary and prevents it from cooling down during the longer timescale run performed here. 
With the above initial setup the whole system now  is out of 
thermal equilibrium. We integrate the governing equations in time 
with heating $H=H_{0}$ active until the system achieves a 
quasi-equilibrium state. After 72 minutes, the above configuration reaches a quasi-equilibrium state shown in Fig.~\ref{3d}, which represents a 3D impression of the numerical box quantifying the temperature and number density profile and selected magnetic field lines. The $t=0$ in Fig.~\ref{3d}  means that after reaching the quasi-equilibrium state, we reset the time of the system back to zero for the next stage of simulation. As seen in this Fig.~\ref{3d}, the numerical relaxation phase leads to some thermodynamic structuring in the final arcade. Some chromospheric plasma is quickly evaporated into coronal loops at the beginning of the relaxation, 
but this material gradually loses its kinetic energy. As a result, the final relaxed state of the system is identified as the time when the maximal residual velocity in the 
simulation is less than 5 km s$^{-1}$. In this end state, Fig.~\ref{3d} shows a relatively 
thin transition region located 
at heights between 3 Mm and 5 Mm, which connects the chromosphere to corona. This transition region is higher above the neutral line, due to less downward thermal flux there because of the strong horizontal magnetic field. The plasma beta is 0.06 at 20 Mm 
height above the neutral line while the temperature and number density there are
around 1.7 MK and $3.5 \times 10^8 {\mathrm{cm}}^{-3}$. The total mass (per unit length in the ignored dimension) of hot plasma in the corona is around $3.2\times10^{4}$ g cm$^{-1}$.

Following this equilibrated system, we turn on a relatively strong heating
$H_{1}$. This extra heating is localized near footpoints in the chromosphere with formula as \citep{fang13}:
\begin{equation}
 H_{1}=\left\{
\begin{array}{lrrrr}
c_{1}  & {\mathrm{if}} & { y<y_{c}} & {\mathrm{and}} & 
{A(x_{1},0)<A(x,y)<A(x_{2},0)}\\
c_{1} \exp(-(y-y_{c})^{2}/\lambda^2) & {\mathrm{if}} & {y\geq y_{c}} & 
{\mathrm{and}} & {A(x_{1},0)<A(x,y)<A(x_{2},0)}
\end{array} \right.
\end{equation}
\begin{equation}
A(x,y)=\frac{B_{0}L_{0}}{\pi}\cos\left(\frac{\pi x}{L_{0}}\right)\exp\left(-\frac{\pi y \sin\theta_{0} }{L_{0}} \right) \,,
\end{equation}
\begin{equation}
\lambda^2=\frac{a\left(A(x,y)-A(x_{2},0)\right)}{A(x_{2},0)-A(x_{1},0)}+b
\rm\quad (Mm^{2}) \,,
\end{equation}
where $c_{1}=10^{-2}$ erg cm$^{-3}$ s$^{-1}$, $y_{c}=3$ Mm, $x_{1}=26$ Mm, $x_{2}=14$ Mm, $a=0.8$ Mm$^{2}$ and $b=1.2$ Mm$^{2}$. This choice of strong base heating contrast ($c_{1}/c_{0}=100$) between $H_{1}$ and $H_{0}$ can mimic flare related chromospheric evaporation.
 However, this $H_{1}$ heating decreases with height to very small values and reaches one-tenth of $H_{0}$ heating at 10 Mm. As a result, the $H_{1}$ heating dominates heating in the chromosphere and the transition region, while the $H_{0}$ heating plays a more important role in the heating in the corona. The parameter $y_{c}=3$ Mm represents the height of the transition region in the quasi-equilibrium system. 
$A(x,y)$ is the magnetic potential depending on the location and decaying exponentially with height into the whole system. Because the magnetic potential along a single magnetic field line is constant, we add extra heating at both feet of all magnetic field lines identified by $A(x,y)$ in the range of $x_{1}<|x|<x_{2}$. Since catastrophic cooling is very sensitive to the heating decay scale and the length of magnetic field lines \citep{xia11}, the heating decay scale $\lambda$ is set to larger values for longer field lines by the above formulae.

\subsection{Discretization, AMR settings and boundary treatment}

We use the MPI-parallelized Adaptive Mesh Refinement (AMR) Versatile Advection Code {\it MPI-AMRVAC}
\citep{amrvac,proth14,keppens14b} to run the simulation. An effective resolution of $4096 \times 2560$ or an equivalent spatial resolution of 20 km in both directions is obtained through six AMR levels. This represents an effective fourfold improvement in resolution with respect to our earlier model \citep{fang13}. Our numerical strategy to advance the governing partial differential equations uses a three-step Runge-Kutta type scheme. For flux computations,  a third-order-accurate limited reconstruction \citep{cada09} is introduced to   calculate the variable evaluation from cell center to cell edge. We adopt a suitably mixed prescription between a diffusive total variation diminishing Lax-Friedrichs and contact-resolving Harten-Lax-van Leer with contact restored (HLLC) scheme \citep{meliani08}.

For boundary treatment, we employ 2 grid layers exterior to the domain as ghost cells to prescribe cell center values. Considering the left and right physical boundary, density, energy, $y$ and $z$ momentum components, $B_{y}$ and $B_{z}$ are set symmetrically, while $v_{x}$ and $B_{x}$ are adopted antisymmetrically to ensure zero face values. In bottom boundary ghost cells, we use the primitive variables ($\rho, \textbf{v}, p, \textbf{B}$) to set all velocity components antisymmetrically to enforce both no-flow-through (vertical) and no-slip (horizontal), while the $\mathbf{B}$ are fixed to the initial analytic expressions of equation~(\ref{bfield}), and the stratification of density is kept at pre-determined values from the initial condition, as well as the pressure. We always resolve the bottom region up to $y = 0.5$ Mm at the maximal resolution. While for the top conditions, we set all velocity components as antisymmetric, and adopt a discrete pressure-density extrapolation from the top layer pressure with a maximal temperature $T_{top} = 2 \times 10^{6}$ K. For the magnetic field, we use a two-cell zero-gradient extrapolation to determine $\textbf{B}$ in the ghost cells and improve $B_{y}$ from a second order one-sided centered difference evaluation of  $\nabla\cdot \mathbf{B}=0$.

\section{Results and Discussion}\label{s-result}
In our previous work \citep{fang13}, we already described the formation process for the first condensation and emphasized how it perturbed the overall force balance in a 2D fashion. In this work, we discuss more of the multi-dimensional details for the forming condensations, and compare the results of our 2.5D simulations with previous 1D simulation works \citep{xia11, muller03} and insights from observations \citep{antolin12}, in particular paying attention to the cross-field effects.

\subsection{Rebound shocks and PCTR of condensations}\label{s-rebound}
\begin{figure}
 \centering
 \includegraphics[width=\textwidth]{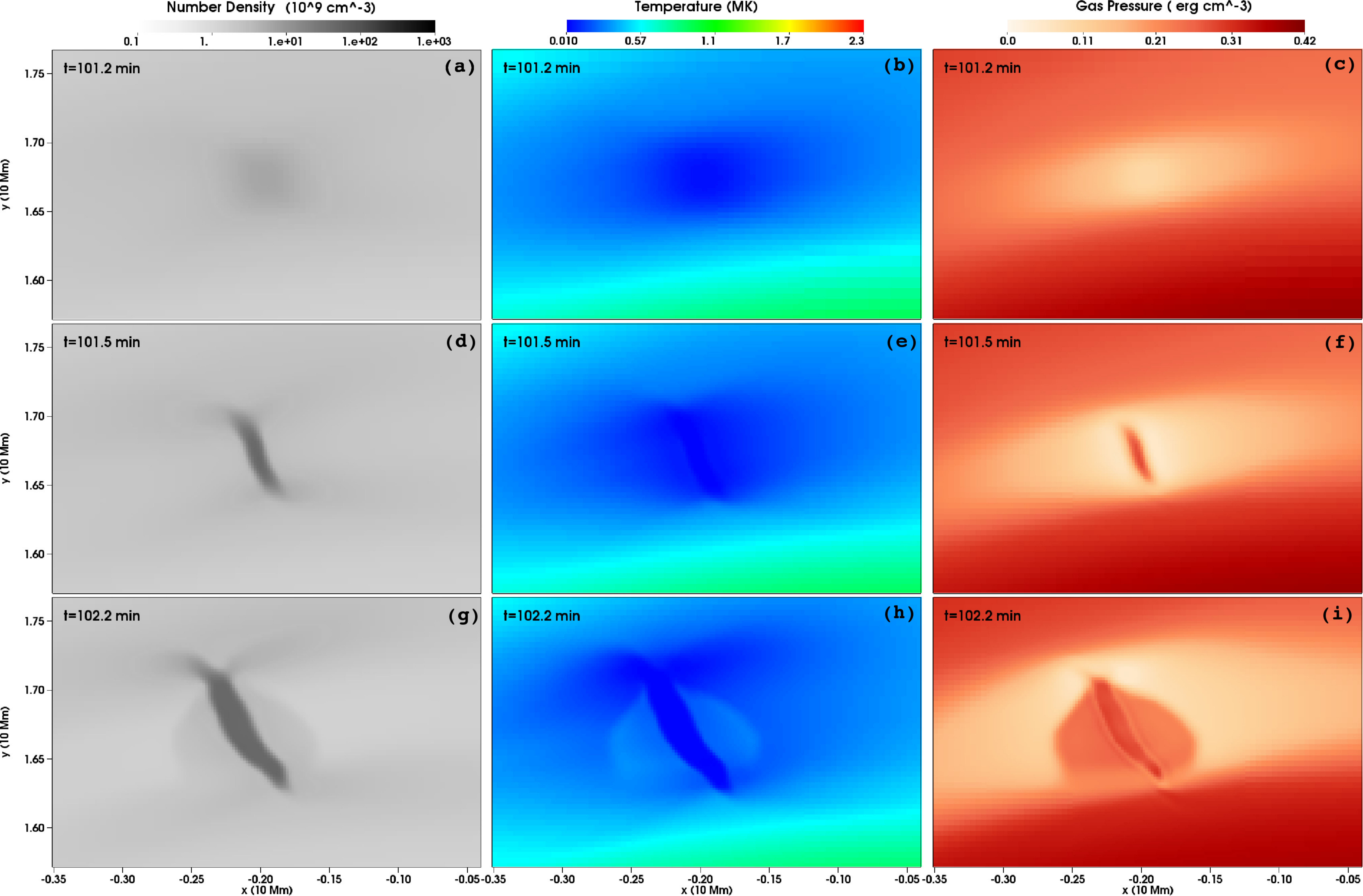}
 \caption{At $t\approx 101.2, 101.5$ and 102.2 minutes (top to bottom rows), we show the number density (left column), temperature (middle column) and
  gas pressure (right column) maps in a zoomed (about $3000\times 2000\,{\mathrm{km}^2}$) area. This shows the formation process of the
  first condensation.}
 \label{first}
\end{figure}

The forming process of the first condensation in our 2.5D simulation is shown by Fig.~\ref{first} which presents the temporal evolution of number density (left columns), temperature (middle columns) and gas pressure (right columns) at $t\approx$ 101.2, 101.5 and 102.2 minutes. When we compare these results with the corresponding Fig. 5 and Fig. 7 of 1D hydrodynamic simulations in \cite{xia11}, we conclude that all three parameters behave similarly in the forming process, as the number density increases rapidly from $10^{8}$ cm$^{-3}$ to $10^{10}$ cm$^{-3}$, while the temperature decreases down to 0.01 MK. Along each arched field line, this is analogous to the sudden thermal instability onset in 1D runs. This similarity confirms the applicability of restricted 1D model efforts which assume a rigid 1D loop under the prevailing plasma $\beta$ conditions, which takes on a local value of around 0.06. The middle panel in the right column of Fig.~\ref{first} also shows a significantly increased gas pressure inside the condensation and a layer of low gas pressure surrounding it after its formation. In the bottom panels of Fig.~\ref{first}, we notice that density, temperature and gas pressure all reveal a front propagating as expanding wings on both sides of the condensation. This phenomenon is because fast siphon inflows are driven into the forming condensation by a strong pressure gradient between the lower gas pressure around the condensation and relatively higher gas pressure away from the condensation, as seen in the middle panel in the right column of Fig.~\ref{first}. These two siphon flows meet up with the blobs, and dynamically impact on the blob to generate two rebound shocks. Hence, while thermal instability and runaway cooling triggers a growing condensation, one also forms two rebound shock fronts that propagate away from the blob. The slightly different formation time at different parts of the condensation on adjacent magnetic field lines \citep{fang13}, which are due to gradual variations in length and chromospheric footpoint conditions, is the reason that these two expansion shock fronts display a fan-shaped structure, forming earliest in the blob center and spreading away from the blob. This fan-shaped structure of the rebound shocks is also clearly observed in \citet{xia12}. 

\begin{figure}
 \centering
 \includegraphics[width=\textwidth]{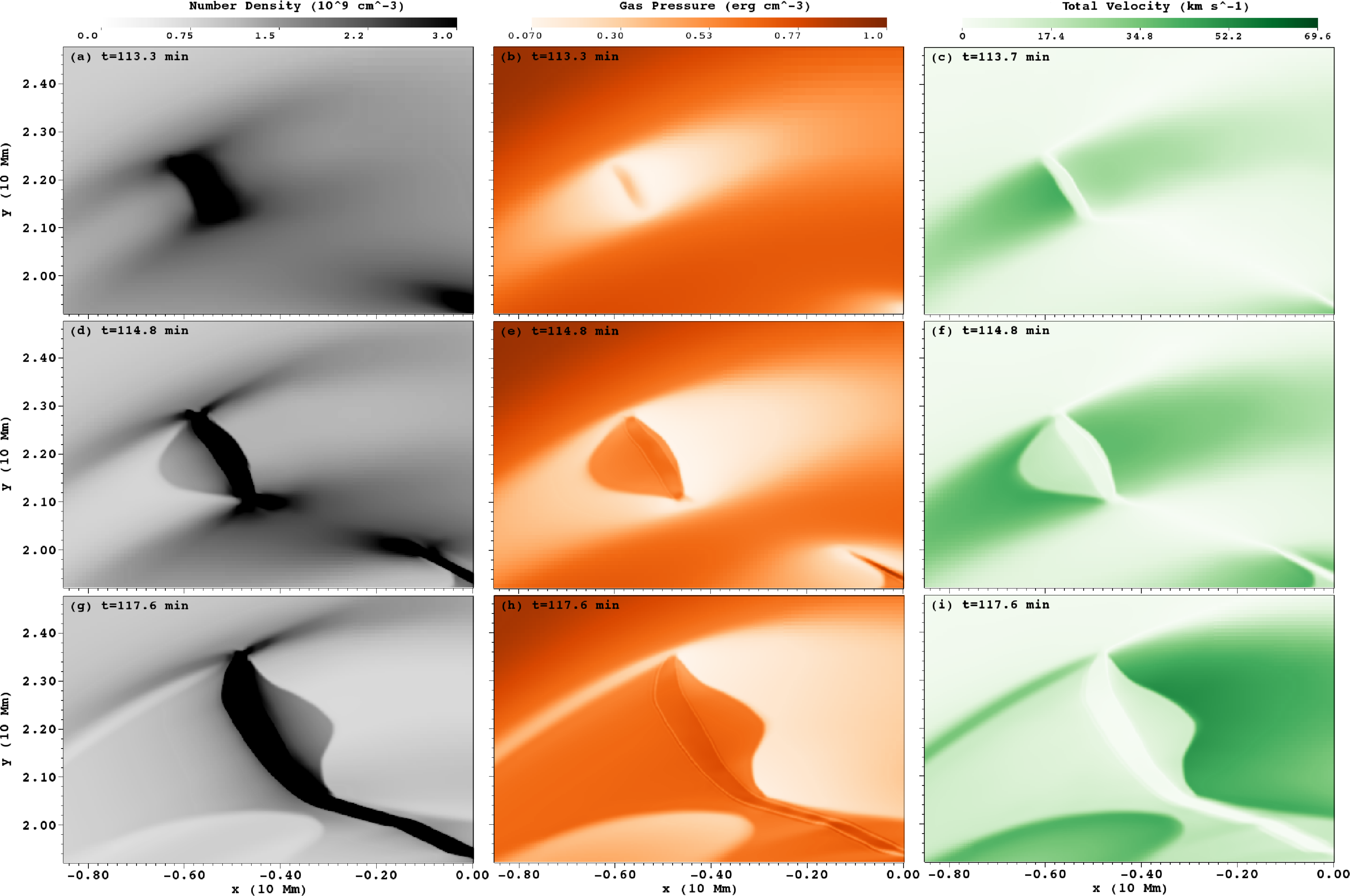}
 \caption{At $t\approx 113.3, 114.8,$ and 117.6 minutes (top to bottom rows), we show evolutions of two-sided rebound shocks in number density (left column), gas pressure (middle column), and plasma velocity magnitude (right column) maps. The panel c shows the plasma velocity magnitude map at $t\approx 113.7$ minutes. This blob shows clear left-right asymmetric behaviour in its rebound shock pair pattern.}
 \label{rs}
\end{figure}

However, not every condensation realizes this nearly left-right symmetric situation as seen near the loop apex for this first condensation from Fig.~\ref{first}. Due to slightly asymmetric conditions already prevailing after the numerical relaxation process and due to perturbations from existing condensations, most of the following condensations initiate in loop limbs (also shown in the online movies of \citealt{fang13}). The field-projected gravity force on the limbs leads to asymmetric plasma distributions, as seen e.g. in the number density map in panel a of Fig.~\ref{rs} at $t\approx113.3$ minutes, the moment when local catastrophic cooling begins there (about 10 minutes after the first condensation). The higher central gas pressure indicates the initial forming location of this condensation in panel b of Fig.~\ref{rs}. Due to its limb-loop location, the number density map points out that the right of the condensation holds a relatively denser ($3\times10^{9}$ cm$^{-3}$) and wider plasma distribution than the left part ($1.5\times10^{9}$ cm$^{-3}$). Still, strong pressure gradients drive siphon flows from both sides towards this condensation. After a short time at $t\approx113.7$, the denser and heavier plasma at the right of this condensation realizes a (left-directed) siphon flow with a slower speed of 23 km/s, compared to the left siphon flow (which is right-directed) at a speed of 42 km/s, shown by the velocity magnitude map in panel c of Fig.~\ref{rs}. As discussed above, the impact of siphon flows on the condensation naturally generates rebound shocks, whose speeds are determined by the original speeds of the siphon flows and the mass contrast between the condensation and the siphon flows. The slower and heavier siphon flow on the right of the blob here leads to a much slower rebound (right-directed) shock seen to separate at 7 km/s, while the left one (left-directed) travels at 21 km/s. These two rebound shocks are identifiable in the gas pressure map in panel e of Fig.~\ref{rs} at $t\approx114.8$ minutes. The condensation itself has a velocity of 5 km/s, meaning that basically the right rebound shock barely can sweep up and heat little siphon flow plasma. Because the central condensation keeps sucking in plasma from nearby and the rebound shock at the right of the blob is too slow to sweep and heat up plasma, the gas pressure there does not rise to a higher value and keeps a strong pressure gradient at the right of the blob, as shown in panel e of Fig.~\ref{rs}. About 3 minutes later at $t\approx117.6$ minutes, this persistent pressure gradient at the right of the blob accelerates the left-directed siphon flow to a higher speed of 52 km/s (shown in panel i of Fig.~\ref{rs}), therefore the corresponding right-directed rebound shock  finally speeds up to 28 km/s and begins to sweep and shock-heat the plasma on its way. In short, initial asymmetric situations on the condensation can lead to a complicated thermal and dynamical evolution and result in a delay of rebound shocks spreading at one side of the condensation. 

\begin{figure}
 \centering
 \includegraphics[width=0.7\textwidth]{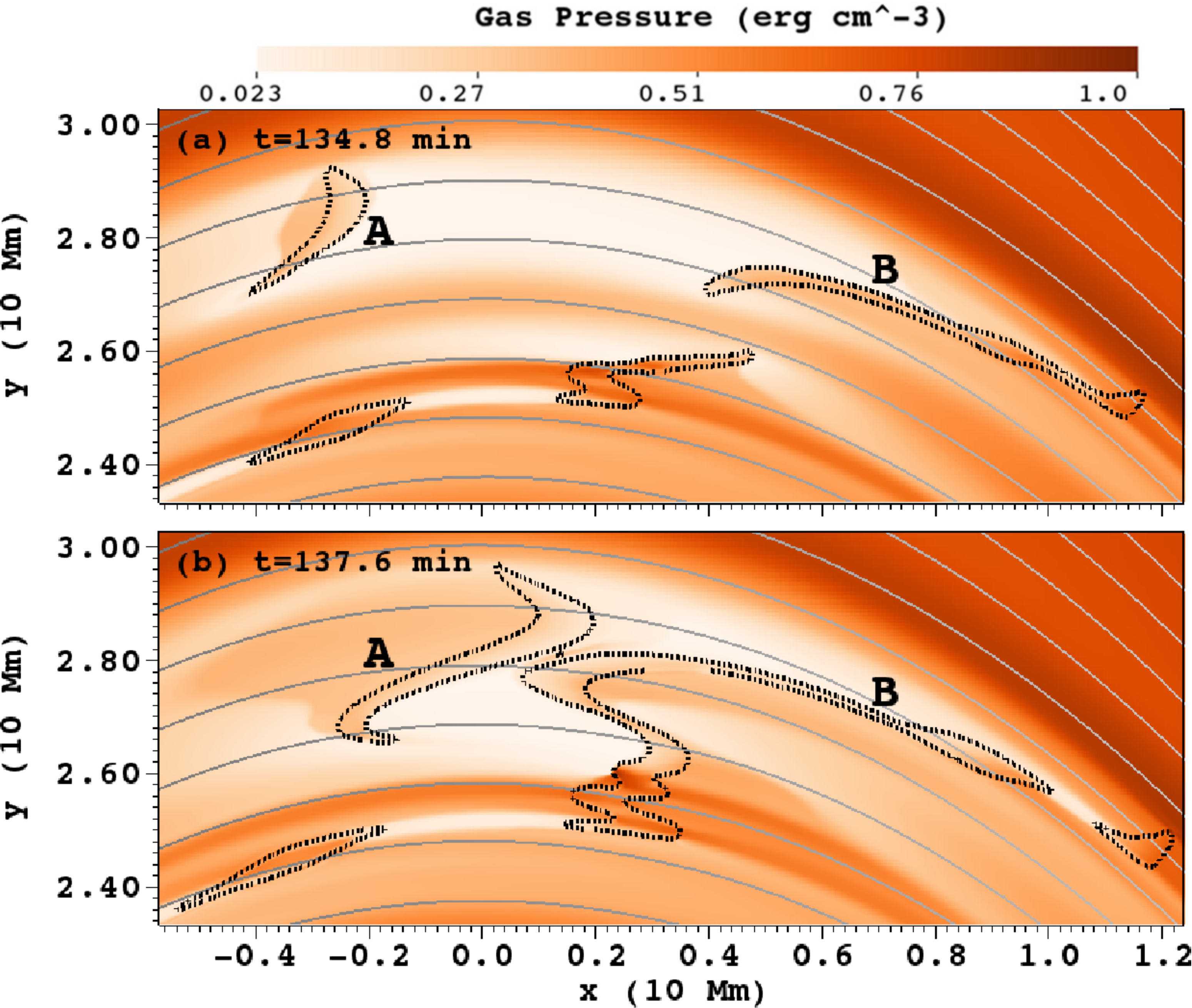}
 \caption{At $t\approx$ 134.8 and 137.6 minutes, we show the gas pressure map (a) and (b) at times indicated, with a dotted isocontour of the number density at $7\times10^{9}$ cm$^{-3}$. The thin grey lines are magnetic field lines. There are two blobs A and B in the same loop, with consequences for the way siphon flows can induce or prevent rebound shock patterns.}
 \label{os}
\end{figure}

Additionally, we also find another special case, namely blob A in Fig.~\ref{os}, which has only one rebound shock on its left side. Fig.~\ref{os}  shows the gas pressure map (a) and (b) at $t\approx$ 134.8 and 137.6 minutes, with a dotted isocontour of the number density at $7\times10^{9}$ cm$^{-3}$ overplotted. This density contour at $7\times10^{9}$ cm$^{-3}$ is one of the criteria which identifies whether a cell contains cool plasma belonging to coronal rain, as used further on. Panel a in Fig.~\ref{os} indicates a similar situation for blob A as in the second row of Fig.~\ref{rs} in which only the left rebound shock spreads out. In contrast to what happens in the third row of Fig.~\ref{rs}, for blob A we do not find a right rebound shock in Fig.~\ref{os} until the collision and merging of blob A with blob B. The reason is that when the thermal instability triggers the condensation labeled there as blob A, another existing condensation labeled as blob B in the same coronal loop already depleted the plasma between these two blobs. Therefore the small pressure gradient in the emptied loop between the two blobs can not drive a fast siphon flow to create a strong rebound shock for blob A, even though the gas pressure on the right of blob A is low enough (panel a in Fig.~\ref{os}),  Afterwards, when blob A catches up and merges with blob B because of the strong pressure gradient outside these two blobs, the rebound shock at the right side of blob A is still not fast enough to show clear separation and propagation. 

\begin{figure}
  \centering
 \includegraphics[width=0.9\textwidth]{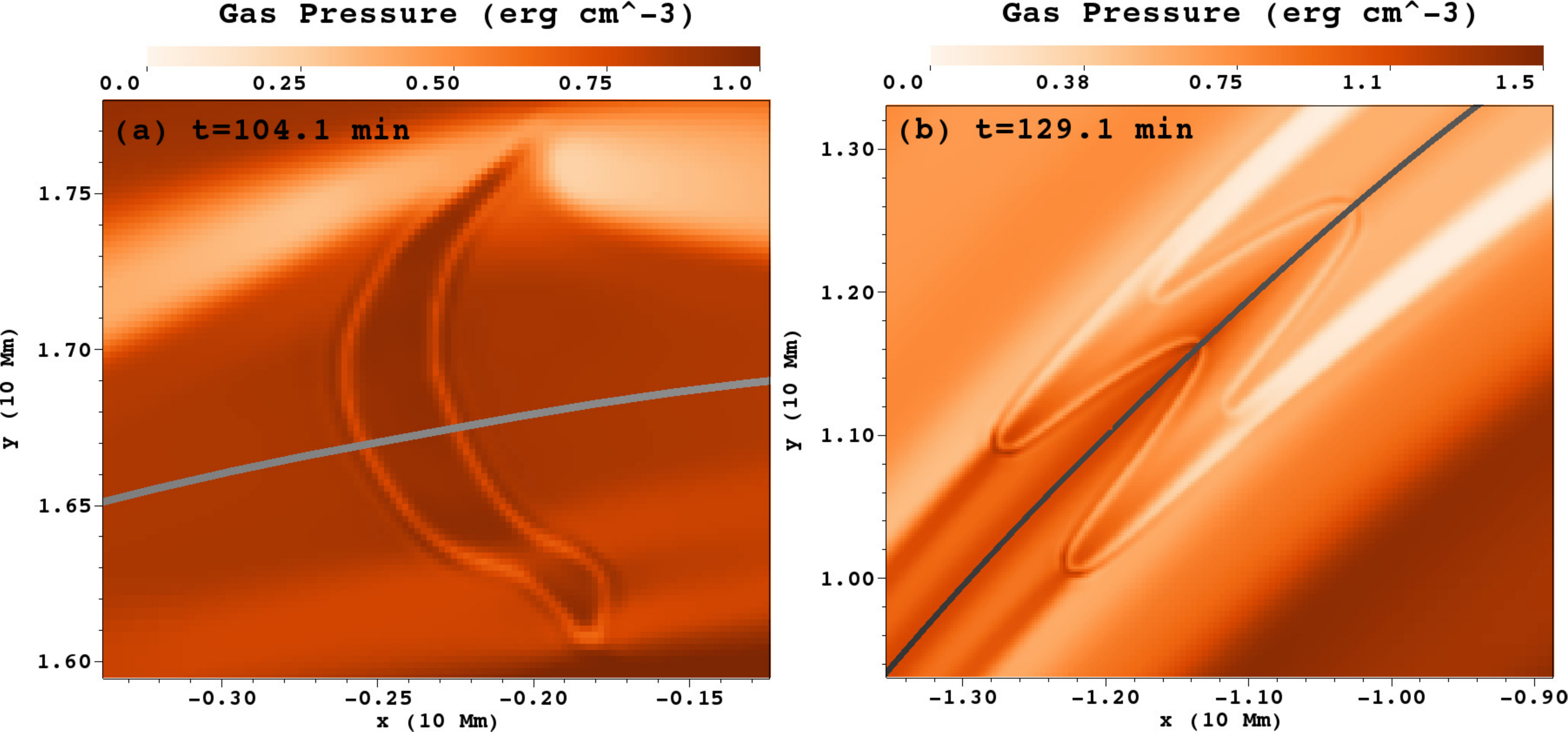}
 \includegraphics[width=6cm,angle=-90]{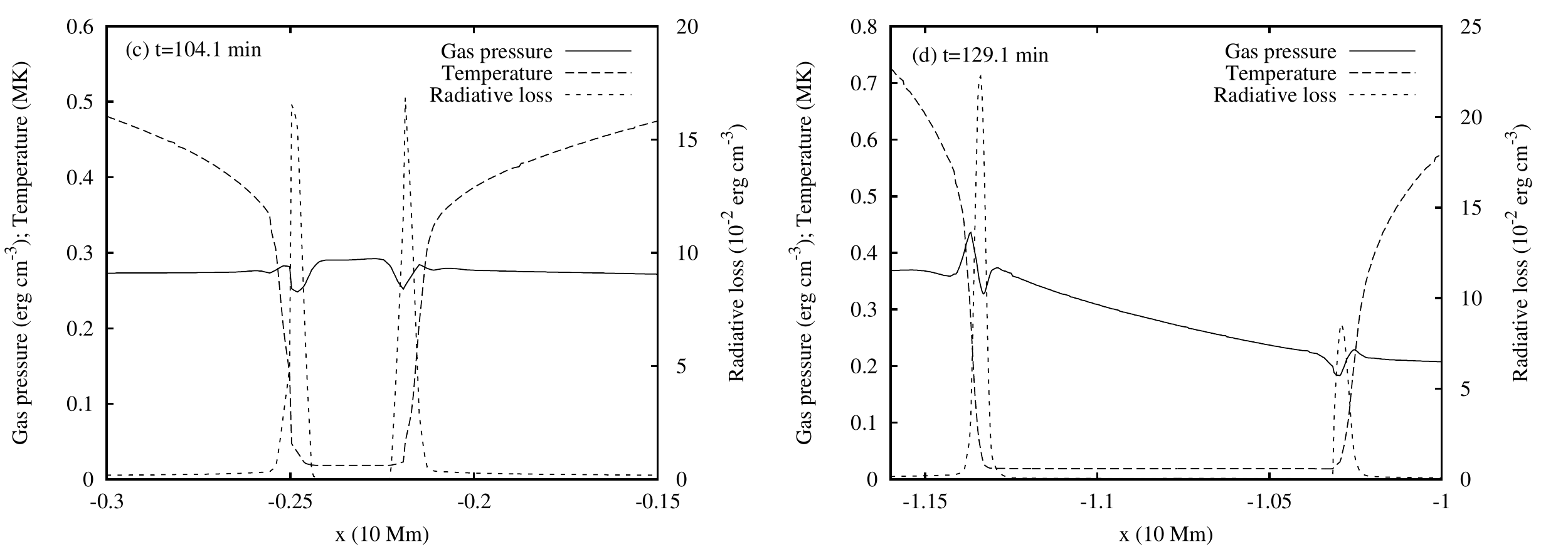}
 \caption{At $t\approx 104.1$ (left) and $129.1$ minutes (right), we show in the top row panels (a) and (b) the gas pressure maps. Panels (c) and (d) plot gas pressure, temperature, and radiation loss along the selected field line crossing the blob center. }
 \label{pctr}
\end{figure}

We also observe the details of a gas pressure substructure within these shock-bounded regions around the condensation in the simulations. These reveal the establishment of a prominence-corona-transition-region (PCTR) like structure around all blobs. The gas pressure substructure around the first condensation consists of three components shown in panel a of Fig.~\ref{pctr} and panel i of Fig.~\ref{first}, namely a high gas pressure outside of the condensation, a low gas pressure at the boundary of the condensation, and a higher gas pressure in the center of the condensation. Actually not only this first condensation in Fig.~\ref{first} has this kind of gas pressure substructure, but also all the blobs which establish a dynamic equilibrium around themselves have it, e.g. all the blobs in Fig.~\ref{rs} and Fig.~\ref{os}. To better quantify this, we identify a field line crossing the center of the blob shown in panel a of Fig.~\ref{pctr} and plot gas pressure, temperature and radiative loss along this field line in panel c of Fig.~\ref{pctr}. The temperature declines from a coronal temperature of 0.35 MK to a cool plasma temperature of 0.01 MK in 200 km and density increases from $1\times10^{8}$ cm$^{-3}$ to $1\times10^{10}$ cm$^{-3}$, therefore basically this 200 km area could be considered as a PCTR. Within this area, we find that two highly radiative loss peaks exist, introduced by a temperature around 0.02 MK. This corresponds to the two dips of gas pressure at the boundary of the blob. These two strong radiation areas also indicate the location in which catastrophic cooling takes place that ensure that the condensation keeps growing. Indeed, the two dips in gas pressure always relocate with the boundary of the blobs, coincident with the strong emissive loss. Although the temperature of 0.01 MK inside the blob is lower than in the surrounding coronal plasma, a much higher density at the center of the condensation ($5\times10^{10}$ cm$^{-3}$) leads to a little higher gas pressure there. The high gas pressure outside of the condensation reflects the post shock conditions prevailing there after the rebound shocks run against the condensation inflows. Note that our resolution is such that we have about 7 grid points along the field line through the PCTR at each side of the blob in Fig.~\ref{pctr}, clearly resolving the PCTR around the blob in our simulation.

The gas pressure difference between inside and outside the condensation is found to persist throughout the lifetime of the blobs and plays a role in the movement of the blobs. Especially when the blobs fall along the field lines toward footpoints, the gas pressure and temperature ahead of the descending blob increase as shown in panel b of Fig.~\ref{pctr}, due to the blob compressing the plasma ahead of it in the loop and the strong evaporation at the loop footpoints. We also identify a field line crossing the center of the blob shown in panel b of Fig.~\ref{pctr} and plot gas pressure, temperature and radiation loss along this field line in panel d of Fig.~\ref{pctr}, which shows also an obvious PCTR. Due to the gravity variation and the strong gas pressure gradient between the two sides of the blob, the lower part of this blob has a higher density distribution, which naturally leads to a higher radiative loss. This strong gas pressure gradient slows down the acceleration of the blob in its descent. This was also pointed out in \citet{fang13}, where we stated that sometimes, it can even lift lighter blobs to cross the loop apex. 

\subsection{Coronal rain limit cycle and condensation rate}

\begin{figure}
 \centering
 \includegraphics[width=6cm,angle=-90]{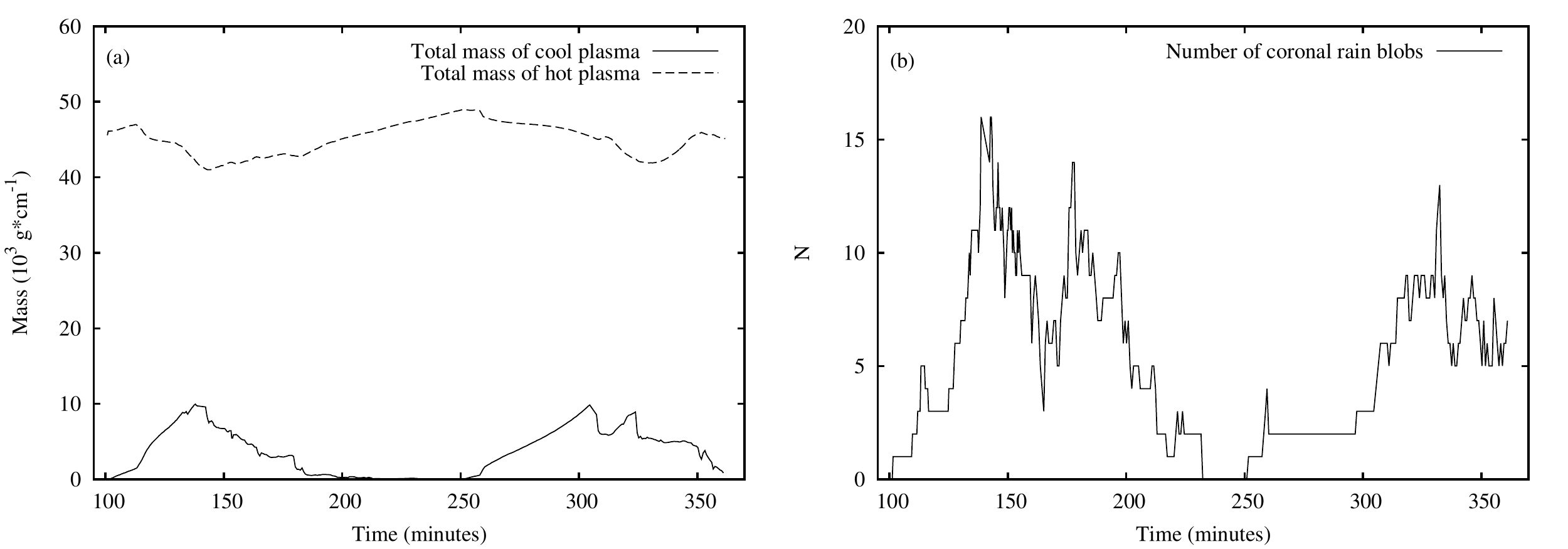}
 \caption{(a) Total mass of cool and of hot plasma in the corona versus time. (b) Number of blobs versus time.}
 \label{tm}
\end{figure}

Panel a of Fig.~\ref{tm} shows the temporal evolution of total mass of cool (solid) and hot (dashed) plasma in the corona, and panel b of Fig.~\ref{tm} represents the number of blobs for the entire time interval of our 2.5D simulation. The criteria to identify whether a cell contains cool plasma belonging to coronal rain are that (i) the number density is higher than $7\times 10^{9}$ cm$^{-3}$, (ii) the temperature is lower than $2 \times 10^4$ K, and (iii) the location is above the chromosphere-corona-transition-region. We dynamically locate the height of the transition region at each $x$-position as $y^{tr}(x,t)$ by searching the vertical position of the (first) maximum gradient value of temperature from the bottom boundary. Each blob is defined as a collection of neighbouring cells which hold cool plasma. However, if the number of grid cells in one blob is smaller than 10 at our highest resolution, we remove this blob from the blob list to avoid counting spurious transient features that do not collect into a clearly resolvable blob, and also to mimic the observational resolution. As stated before, we adopt a 4 times higher resolution than in \citet{fang13}, but also extend the simulation to a two times longer time of around 370 minutes (previously 190 mins). By running our 2.5D simulation for these much longer times, we find that the whole coronal rain process shows limit cycles, which has been discussed in earlier 1D simulations \citep{muller03}, as well as in observational work \citep{antolin12}. This is the first time that we can report limit cycles of coronal rain in a multidimensional simulation, which confirms that constant heating conditions which provide enough energy, can form secondary (or even more) coronal rain cycles in a single arcade. From panel a and b of Fig.~\ref{tm}, we find the time interval between the first and secondary cycle to be around 175 minutes, when measured between successive maxima in cool mass matter.
Panel a of Fig.~\ref{tm} shows the temporal evolution of total mass of hot coronal plasma which is the mass in the corona, excluding the cool plasma identified by the above criteria. We find that at $t\approx140$ minutes, the total mass of cool plasma reaches its peak in the first cycle in panel a of Fig.~\ref{tm}, while at $t\approx143$ minutes the catastrophic cooling process has cooled down most of the hot plasma in the corona shown in panel a of Fig.~\ref{tm}. From about $t\approx133$ minutes, blobs begin to fall into the transition region, then the evaporation of plasma in the chromosphere driven by the extra heating $H_{1}$ fills the evacuated loops left by blobs which already sank into the chromosphere. From this moment, until the onset of the secondary cycle of our coronal rain shower at $t\approx250$ minutes, it takes about 120 minutes, which is of similar duration as the time for the first cycle to reach its onset (about 100 minutes). So although we infer from the total mass evolution of cool plasma in panel a that there is only about 50 minutes between the ending of the first and the beginning of the second cycle, actually the continued heating at the chromosphere spontaneously begins to fill the empty corona already 70 minutes before. We also see that the total mass of hot plasma before the onset of the secondary cycle is higher than in the first cycle (panel a), which leads to a longer lasting secondary cycle with more mass in condensations. Panel a of Fig.~\ref{tm} indicates that at $t\approx130$ minutes (before the first blob falls into the chromosphere), there is at least $9\times10^{3}$ g cm$^{-1}$ cool plasma in the corona, which originally was hot plasma. Meanwhile panel a of Fig.~\ref{tm} also suggests that compared with the corona before the onset of catastrophic cooling at $t\approx100$ minutes, the decrement in the same time of total mass of hot plasma at $t\approx140$ minutes is only $5\times10^{3}$ g cm$^{-1}$. The difference between the increase in cool plasma and the decrease in hot, indicates that during these 30 minutes since onset at $t\approx100$ minutes, the evaporation in the chromosphere evaporates $4\times10^{3}$ g cm$^{-1}$ into the corona, i.e. at an evaporation rate of 2.2 g cm$^{-1}$ s$^{-1}$. We can similarly estimate an evaporation rate of 2.3 g cm$^{-1}$ s$^{-1}$ between the onset of the secondary cycle and the moment its first blob falls into the transition region. 
Till the onset of the first cycle at $t\approx100$ minutes, the increment of total hot plasma from turning on the extra heating $H_{1}$ is about $13.2\times10^{3}$ g cm$^{-1}$ in total, further confirming this evaporation rate of 2.2 g cm$^{-1}$ s$^{-1}$. Based on these estimates, we infer that anywhere in both simulated cycles, the constant extra heating $H_{1}$ leads to a nearly constant evaporation rate. We can thus extrapolate to even more cycles expected further on, and interpret these limit cycles as a chronological sequence of mass recycling between chromosphere and corona: heating in the chromosphere brings plasma to the corona by evaporation, where it ultimately triggers catastrophic cooling, the cooling process manages itself into a coronal rain where plasma drains back to the chromosphere, and persistent heating causes the chromospheric material to evaporate again towards the corona. 

Although the duration and peak value of the total mass in both computed cycles are similar, their initial condensation rates (in contrast to the previously discussed evaporation rate) computed from the temporal variation of their total mass curve work out to be 6.7 and 4.5 g cm$^{-1}$ s$^{-1}$, respectively and thus are different. It is known from linear thermal instability theory \citep{field65} and 1D simulation results in \citet{xia11}, that this initial condensation rate in catastrophic cooling depends on parameters controlling the energy input from heating. One notices that the condensation rate (the local derivative of the solid curve in panel a of Fig.~\ref{tm}) varies dramatically even within one cycle, despite a constant heating energy input in our multidimensional simulation. We now will interpret the reason for the changes seen in the condensation rate, by surveying especially the process of growth for the first condensation which forms under a relatively simple and almost symmetric condition.
 
\begin{figure}
 \centering
 \includegraphics[width=11cm,angle=-90]{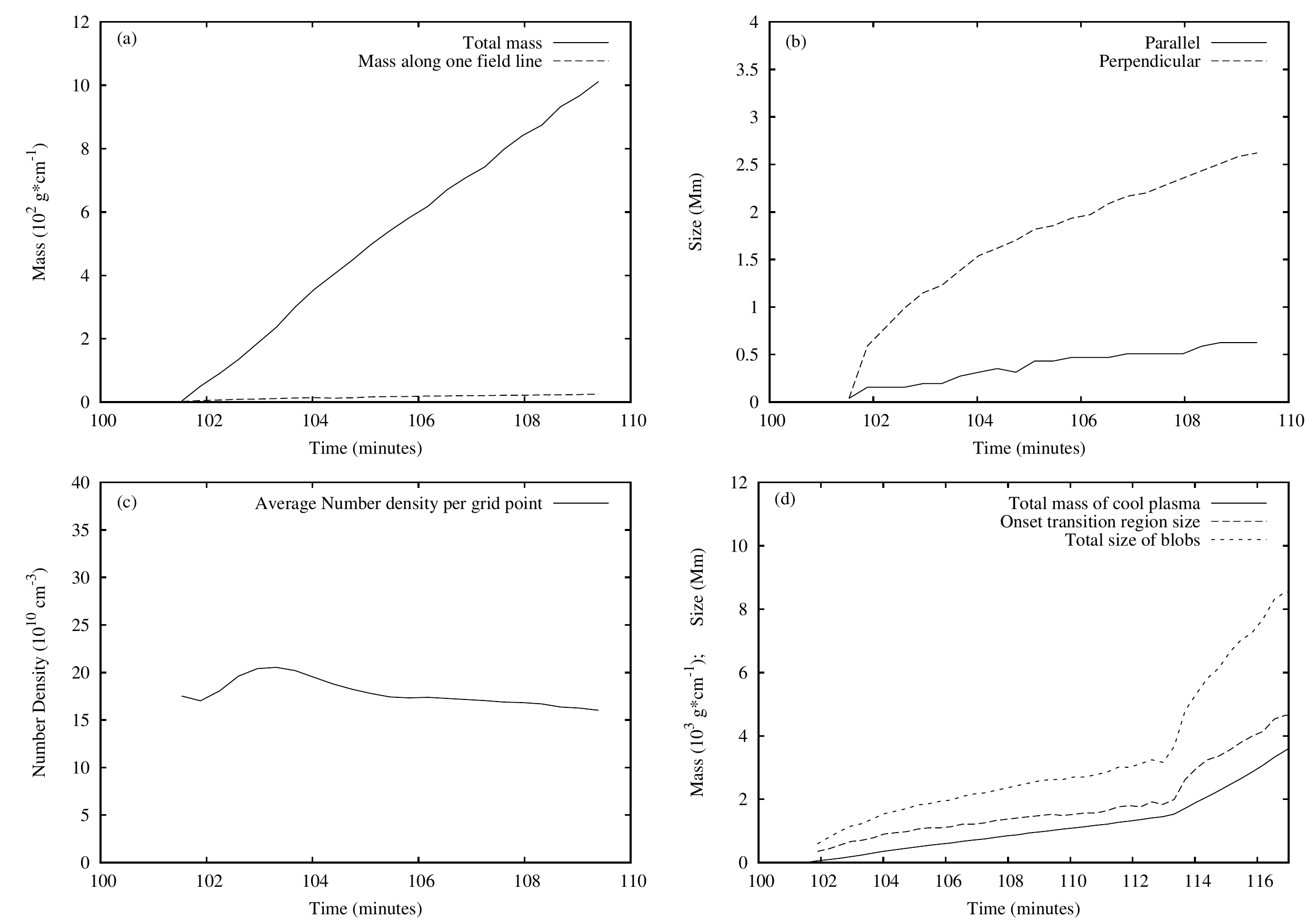}
 \caption{(a) Mass in the first condensation versus time. The dashed line shows a measurement performed along one field line only.
  (b) Lengths versus time of the first condensation (solid: length parallel to the magnetic field lines; dashed: length perpendicular to the magnetic field lines).
  (c) Average density evolution of the first condensation.
  (d)Total mass of cool plasma in the corona (solid), onset transition region size (long dashed) and total size of blobs (short dashed) versus time within the range from 100 minutes to 117 minutes.}
 \label{first_t}
\end{figure}

The solid line in panel a of Fig.~\ref{first_t} shows the temporal evolution of the mass accumulation for this first condensation (the one from Fig.~\ref{first}) from $t\approx100$ to 110 minutes. Its near linear behavior quantifies that the condensation rate remains almost constant in this time interval at a value of about 2.3 g cm$^{-1}$ s$^{-1}$. 
We deliberately do not discuss what happens to the first condensation after $t\approx110 $ minutes, since afterwards it breaks into two smaller blobs. 
In the same figure panel a, the dashed line displays the growth of the total mass of cool plasma as seen on a single field line through the center of the first condensation, i.e. in a 1D fashion. To show this, we identify the group of grid points which are passed by the field line.
The total mass of cool plasma determined on the single field line keeps growing in time, but its growth rate is much smaller than that for the whole 2D condensation. Panel b of Fig.~\ref{first_t} quantifies the temporal evolution of typical lengths for the first condensation, where we quantify both the length parallel to the magnetic field lines and the length perpendicular to the magnetic field lines. This indicates that blob growth in the perpendicular direction is much faster than in the parallel one, which can be seen visually as well in all columns in Fig.~\ref{first} and Fig.~\ref{rs}. As discussed in \citet{fang13}, the low pressure region surrounding the first condensation onset leads to magnetic restoring forces on adjacent loops. These inturn influence in which location the catastrophic cooling will take place on the adjacent loops, which are all close to the thermal instability onset. The different growth rates found for blob sizes in these two directions then relate to the fairly fast `growth' along the perpendicular direction due to sympathetic runaway cooling onset, versus the slower growth seen in the parallel one, which is the only one found in 1D setups. The average density of each cell of the condensation is quantified in panel c of Fig.~\ref{first_t}, and this density stays basically the same in the forming process, meaning that the total mass of the condensation is just proportional to the increasing number of neighbouring grid cells that contain cool plasma. While the number of cells in the condensation increases in both directions, the larger condensation rate of the whole blob in panel a of Fig.~\ref{first_t} again directly reflects the faster growth in size in the perpendicular direction. We conclude that the growth of total mass of individual blobs in our simulation is mainly determined by the onset of catastrophic cooling in neighbouring loops rather than the growth along the loops in which catastrophic cooling gets triggered. 
We can indeed verify this 2D growth aspect by further showing a correlation between the total mass of cool plasma and two other measures, which holds up even for a longer time than the first 10 minutes, i.e. when several blobs have started to form. This is shown in panel d in Fig.~\ref{first_t} where we plot the temporal evolution of total mass of cool plasma, the size of the onset transition region, and the total blob region width. The total blob region width indicates the total width of all magnetic loops where catastrophic cooling takes place on. The size of onset transition region means the corresponding width as found at the transition region height, of all the loops undergoing catastrophic cooling. Because the magnetic arcade configuration adopted, these size measures for the affected loops give higher values for higher locations, i.e. the total blob region width always exceeds the (field aligned remapped) onset transition region size. The latter size of the onset transition region shows a nice correlation with the total mass of cool plasma evolution. 

\subsection{The fate of blobs hitting the transition region}\label{s-impact}

\begin{figure}
 \centering
 \includegraphics[width=15cm]{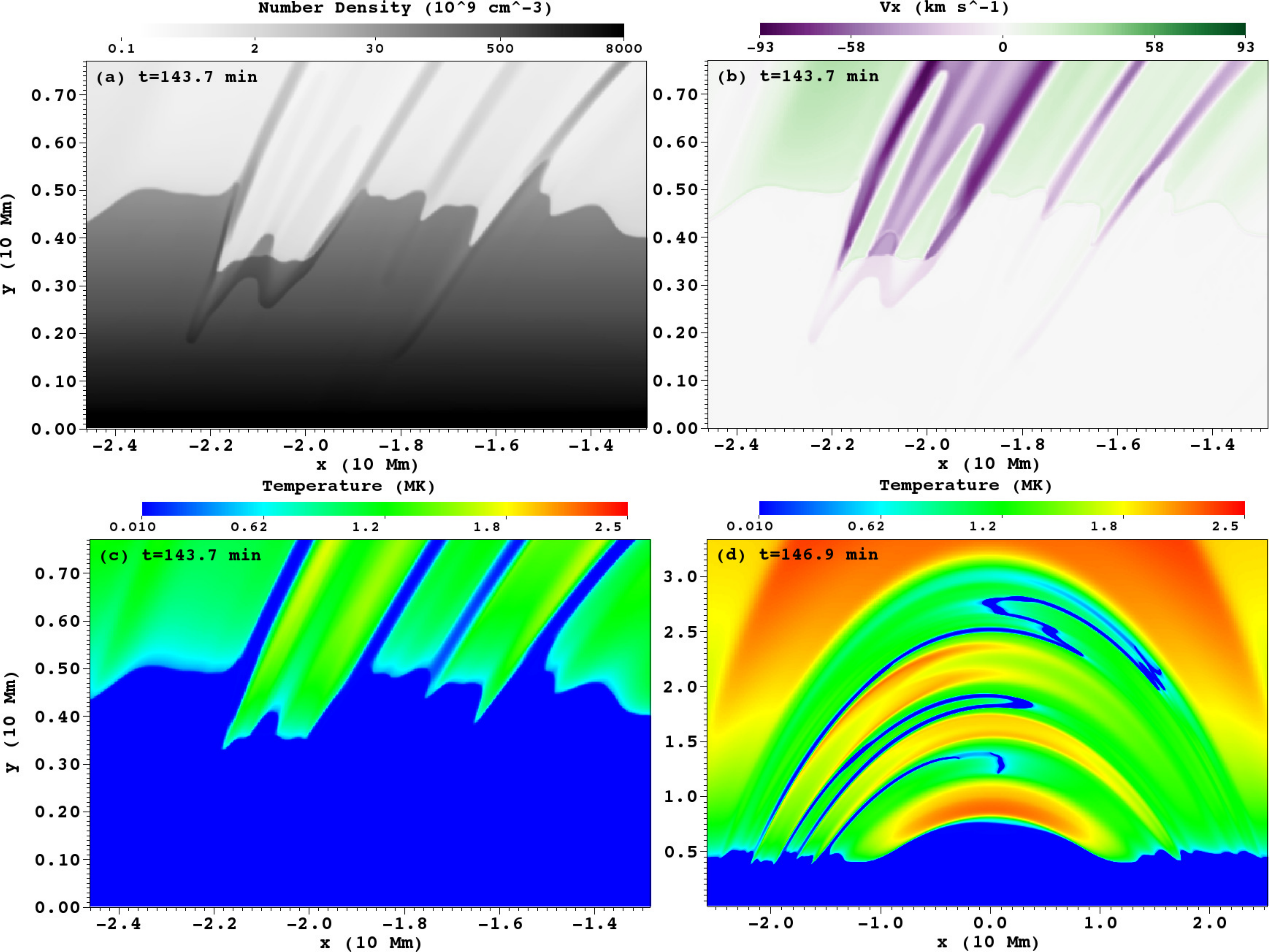}
 \caption{(a) The number density map at $t\approx143.7$ min; (b) The horizontal velocity component map; (c) The temperature map. These three panels show the same local area with chromosphere and transition region variations, while a larger area view is shown in panel (d), giving a later temperature map at $t\approx146.9$ min.}
 \label{hit1}
\end{figure}

In the simulation, we observe plenty of blobs hitting the transition region and disappearing into the lower chromosphere, as also known to occur in observations \citep{antolin10, antolin12}. \citet{tripathi09} observed high-speed downflows and concurrent upflows in coronal loops close to the footpoints and argue in favor of upflows in coronal loops at higher temperatures. \citet{antolin10} confirmed that the high-speed downflows represent the cool plasma, which is corresponding to the falling blobs in our simulation (see once more also the online movie of \citealt{fang13}). Meanwhile, our 2.5D simulation also shows the possibility of triggering concurrent upflows as observed by \citet{tripathi09} and \citet{kleint14}. Panel a in Fig.~\ref{hit1} shows the number density map at $t\approx143.7$ at a moment when falling blobs sink into the transition region and compress the plasma on its way (at about $x\approx -2.1$ Mm). Panel b in Fig.~\ref{hit1} shows the vertical velocity map at the same location and instant, which clearly displays the concurrent upflows rising at the tails of the declining blobs in the same field line bundles. Hence, this answers the question in \citet{kleint14} whether the upflows can flow along the same field lines as the downflows. These upflows in our simulation are actually rebound shocks from the impact of the blobs on the transition region (TR). They arise immediately when the blobs impact on the TR, and spread from one footpoint to another footpoint in around 5 minutes with a velocity of around 50 km s$^{-1}$ . From Panel a we can see the enhanced density left after passage of these rebound shocks. However, panel c in Fig.~\ref{hit1} which quantifies temperature indicates that the temperature in the loop already increases before the rebound shocks have reached far into the loop, since the parametrized background heating $H_0$ heats the low density loops left by falling blobs very efficiently. Panel d shows also the temperature, but now on a larger domain and at a later time. It shows that afterwards the rebound shocks heat the low density loops to an even higher temperature of 2.0 MK. After the rebound shocks reach the other footpoint, the loops are at high temperature of about 2 MK but with a low number density of $1\times10^{8}$ cm$^{-3}$. We distinguish this from further upflows coming from evaporation due to the extra strong heating $H_{1}$ located in the chromosphere. This enhances the density to $1\times10^{9}$ cm$^{-3}$ again and the temperature to 2.3 MK. However, these upflows from evaporation rise with a much slower velocity of around 15 km s$^{-1}$.

\begin{figure}
 \centering
 \includegraphics[width=11cm,angle=-90]{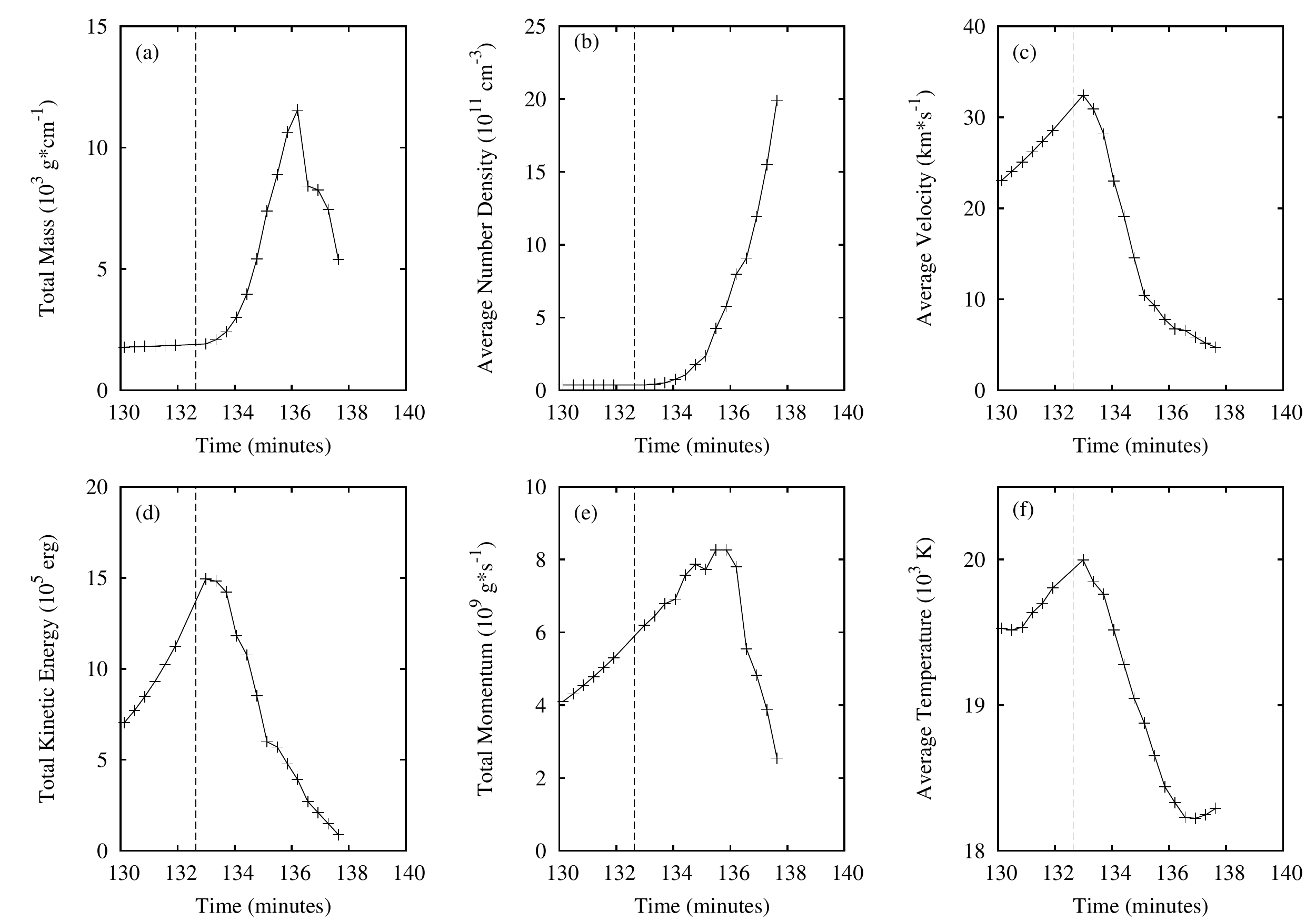}
 \caption{We show the total mass, average number density, average velocity, total kinetic energy, total momentum and temperature evolution of the blob which first impacts and sinks into transition region, during this time period. Vertical dashed line indicates the time when this blob hits the transition region.}
 \label{hit}
\end{figure}

To quantify even further the detailed fate of a blob when it hits and descends into the TR, Fig.~\ref{hit} shows the temporal evolution of the mass, density, velocity, kinetic energy, momentum, and temperature of the first coronal rain blob to hit the transition region from the corona and to sink down into the chromosphere. 
The vertical dashed line in each panel of Fig.~\ref{hit} points at $t\approx132$ minutes when this blob hits the transition region. Because the density and temperature of plasma in the transition region is comparable with those of the blobs, we can not use only the density and temperature as a criterium to distinguish blobs when they are near or partially below the transition region anymore. In order to identify plasma belonging to the blob as it hits and descends in the transition region after $t\approx132$ minutes, we change our criteria to require the local velocity to be larger than 3 km s$^{-1}$ and the location are below the transition region line $y^{tr}(x,t)$ after $t\approx132$ minutes.  Since the velocity of plasma in the transition region is almost zero, this velocity-based criterion captures the location of sinking blobs. 
In panel a of Fig.~\ref{hit}, we find that the mass identified as blob material by the above criteria begins to increase at t$\approx$ 132.7 minutes. This is because the mass detected not only includes the blob itself, but also counts mass compressed and accelerated by the blob impact. At $t\approx 136$ minutes, the total mass affected reaches its peak at six times the original blob mass. After $t\approx 136$ minutes, due to the combined influence of reflection-transmission processes at the transition region, and the higher gas pressure from the impact, the velocities in much of the blob impacted area decrease to values smaller than the criterion 3 km s$^{-1}$. This is then seen as a mass decrease in our panel a. In panel b of Fig.~\ref{hit}, the density versus time profile keeps rising while the blob hits the transition region. As we know, this blob impact compresses the transition region plasma swept up by the blob and transfers momentum from the sinking blob to the impacted plasma, and therefore in panel c of Fig.~\ref{hit} we find that the average velocity of the region identified keeps decreasing during the whole process, as well as the kinetic energy shown in panel d. Panel e of Fig.~\ref{hit} shows the total momentum of the mass identified. Due to the gravitational acceleration, the blob momentum keeps increasing until it reaches its maximum value at $t\approx$ 136 minutes, then it reduces quickly. This is a combination of the mass evolution in panel a and the velocity info from panel c. The momentum and velocity decreasing after the impact relate to momentum transfer to the surrounding transition region and upper chromosphere plasma, until the regions selected by the velocity-based criteria vanishes: the local conditions settle to static chromosphere conditions. Panel f of Fig.~\ref{hit} shows the average temperature evolution during the blob impact. The temperature increases before hitting the TR due to the compressional heating when the blob descends through the higher gas pressure region just above the transition region. After the impact, since also more cooler material gets identified as impacted matter, one settles back to upper chromospheric temperature values. 

The impact speed of blobs in Fig.~\ref{hit} is around 30 km s$^{-1}$, and the highest impact speed of all blobs in our simulation is around 60 km s$^{-1}$ and number densities range from 4 to 6$\times10^{10}$ cm$^{-3}$. Our maximum impact speed is much lower than the falling speeds reported in \citet{kleint14} which went up to 200 km s$^{-1}$. They report that these coronal rain events with high impact speeds are correlated with local brightenings which probably indicate an increase of density and temperature in the transition region. Panel b of Fig.~\ref{hit} and panel a of Fig.~\ref{hit1} confirm the expected increase of the number density of impacts in our simulation.

\subsection{Counter-streaming flows}

\begin{figure}
 \centering
 \includegraphics[width=15cm]{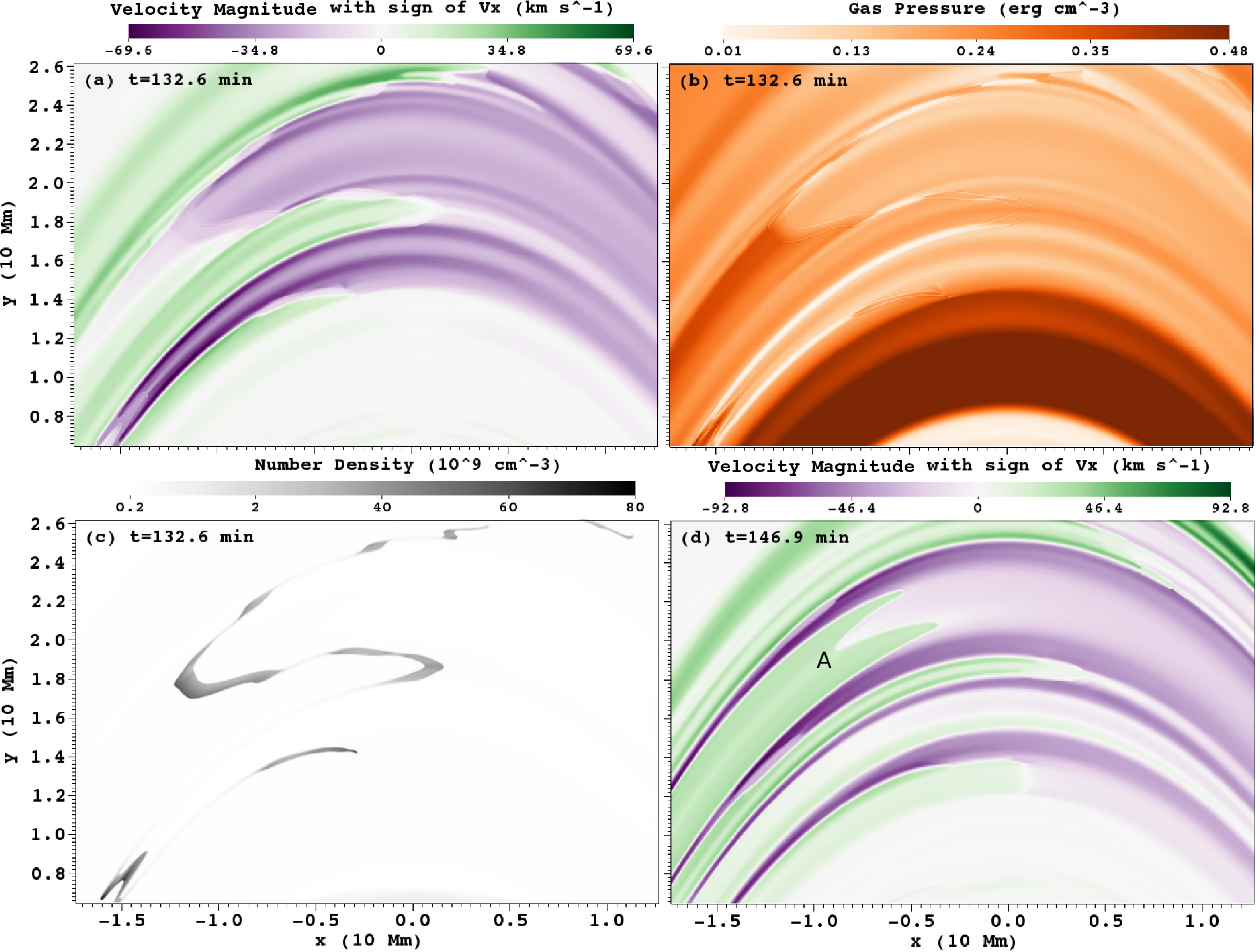}
 \caption{(a) shows the velocity magnitude map signed with horizontal velocity component at $t\approx132.6$ minutes. Panel (b) shows the gas pressure map and panel (c) shows the number density map at the same time. In panel (d), the signed velocity magnitude map is shown later at $t\approx146.9$ min, where the label A points to the upflows resulting from the rebound event shown in detail in Fig.~\ref{hit1}.}
 \label{sf}
\end{figure}

\begin{figure}
 \centering
 \includegraphics[width=19cm]{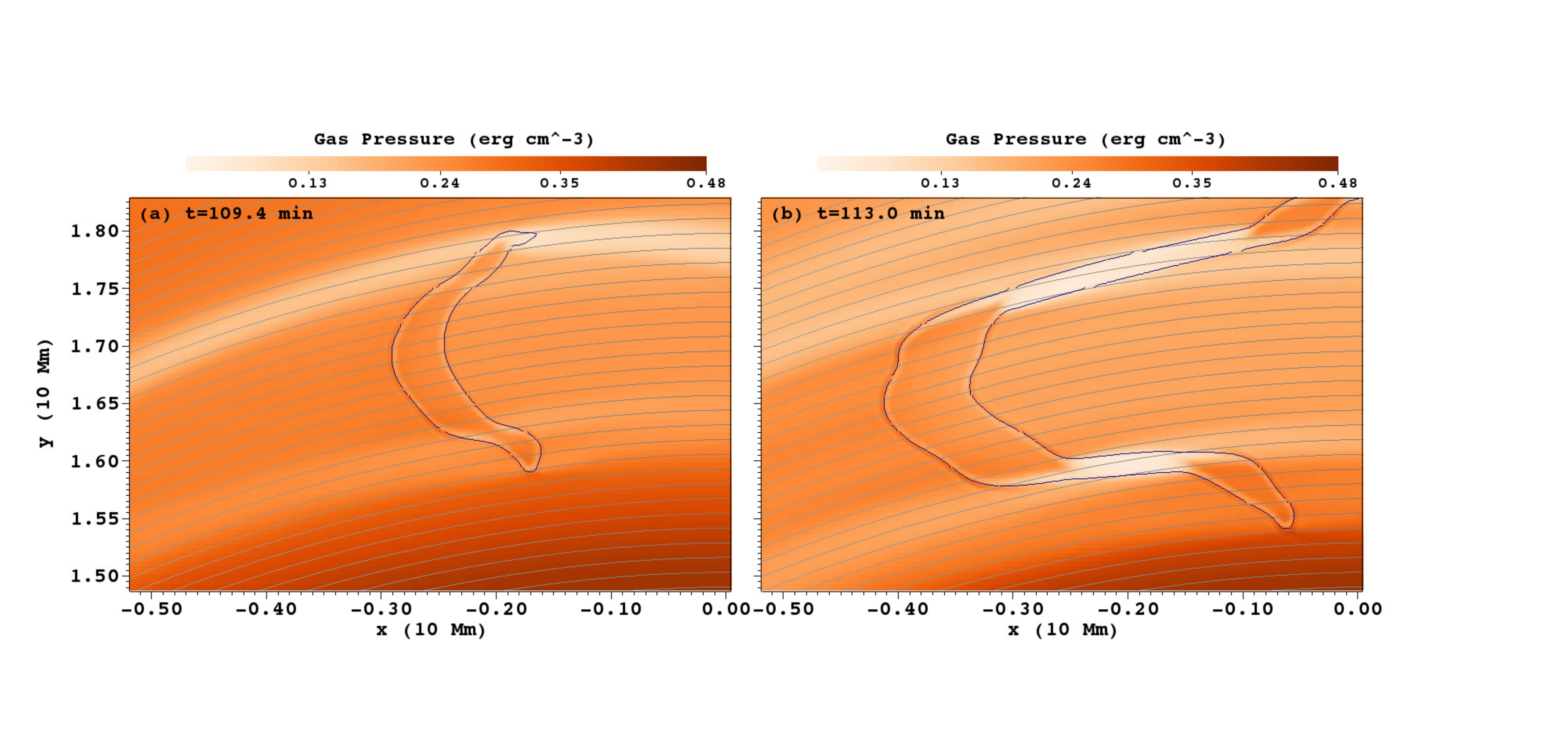}
 \caption{Panel a and b show the gas pressure maps with magnetic field lines at  $t\approx109.4$ and 113.0 minutes. The black contour relates to the temperature distribution with level at 0.1 MK.}
 \label{lp}
\end{figure}

We also find another interesting phenomenon in our numerical simulation, namely the self-consistent establishment of counter-streaming flows. Such anti-parallel flows are very commonly found in solar observations, especially also in prominences \citep{alexander13}. Panels a, b and c in Fig.~\ref{sf} respectively show the signed velocity magnitude map (with the sign taken from the horizontal velocity component), the gas pressure map and the number density map at $t\approx132.6$ minutes. Panel d shows the signed velocity magnitude map as in panel a, but at a later time, namely at $t\approx146.9$ minutes. These four panels in Fig.~\ref{sf} display many cases of counter-streaming flows established on neighbouring field line bundles in our simulation and allow to explain the origins of counter-streaming flows. After thermal instability inducing a runaway catastrophic cooling and initial growth in an almost static state, the condensations lose their delicate force balance and begin to slide towards one footpoint along magnetic field lines. Whether a particular condensation segment slides to the left or right is influenced by its initial location and local total force balance (gravity, gas pressure gradient and magnetic field force). Once in motion, they are accelerated by the field-projected gravitational force, meanwhile catastrophic cooling keeps taking place around the condensations. As discussed in Section~\ref{s-rebound}, the initial catastrophic cooling process depletes the local plasma and sucks in fast inflows, then the spontaneous rebound shocks heat the plasma and increase the gas pressure. Afterwards, no stronger inflows are driven again due to the increased gas pressure. However, there can be several blob pairs lying in the same or neighbouring field line bundles, as shown in panel c of Fig.~\ref{sf}. Fig.~\ref{lp} shows gas pressure maps with magnetic field lines at  $t\approx109.4$ and 113.0 minutes. The black contour relates to the temperature distribution and is an isocontour at 0.1 MK. Both the gas pressure and temperature in panel a in Fig.~\ref{lp} indicate the clear PCTR around the blob as previously discussed in Fig.~\ref{pctr}. After 3 minutes, the panel b of Fig.~\ref{lp} shows two white (low) pressure sections after the blob breaks into three segments. These low pressure sections slant through the field lines and they are the elongated PCTR cross sections from the original parts of the whole blob in panel a of Fig.~\ref{lp}. Because the strong radiation in the PCTR, the temperatures inside these elongated regions stay low during their deformation. As a result, we could consider these cross sections to undergo an isothermal expansion. Because the condensed mass in these narrow regions grows much slower than their areal growth due to elongation, the densities inside these elongated cross sections decrease faster as well as the gas pressure. This leads to blob sequences with low pressure sections in between them. This is also seen in panel b of Fig.~\ref{sf} where a sequence of blobs show up with white (low) pressure sections in between them. The depleted areas trigger siphon inflows to refill these regions. Then this pair of siphoned fast inflows establish the counter-streaming flows between the pair of neighbouring blobs. Panels a and b in Fig.~\ref{sf} also show that falling condensations with larger velocities induce larger density depletions and lower gas pressure areas on their way down, which leads to faster inflows than those found for more static condensations. 

Panel d in Fig.~\ref{sf} indicates another different origin of counter-streaming flows at $t\approx146.9$ minutes. As we discuss in Section~\ref{s-impact}, we observe that after blobs decline into the transition region, concurrent upflows rise up towards the loop apex. Upflows labeled as A in panel d in Fig.~\ref{sf} are the concurrent upflows shown in panel b of Fig.~\ref{hit1}, but about 3 minutes later (concurrent with the later temperature panel d of Fig.~\ref{hit1}). Upflows arising from blob impacts also have the chance to establish a counter-streaming flow if there is an opposite flow pattern in the neighbouring loops. The difference between these two different origins for counter-streaming flows is that the one based on depleted sections between a pair of blobs lying on neighbouring field line bundles can last through the whole falling process of blobs, or on time scales of about 10 minutes, while the other ones will vanish after the upflows refill the loops, typically in a shorter time scale of about 5 minutes. 

\subsection{Shear flow effects}
\begin{figure}
 \centering
 \includegraphics[width=15cm]{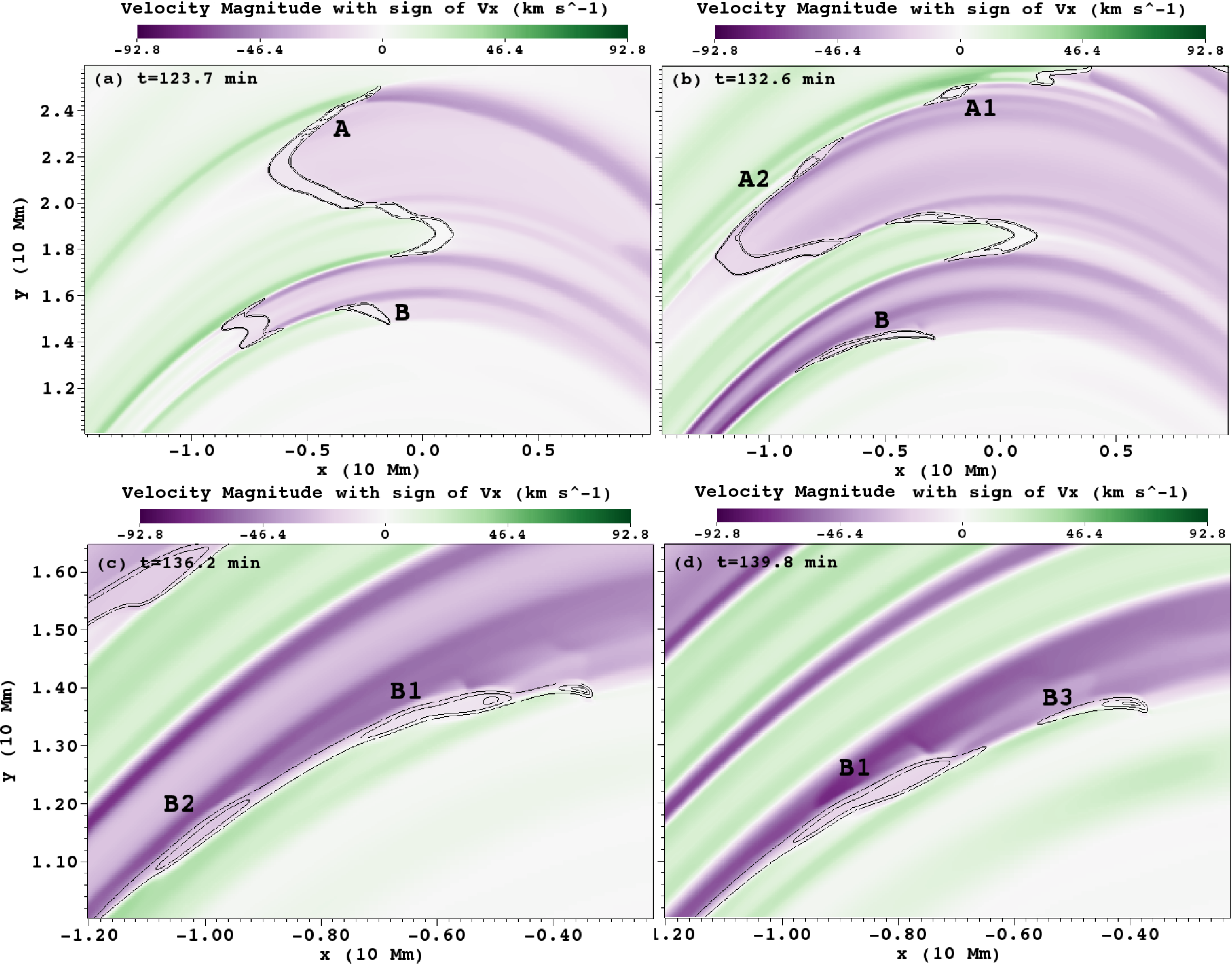}
 \caption{Color maps show the velocity magnitude map with the sign of the horizontal velocity component at $t\approx123.7$ (a), 132.6 (b), 136.2 (c) and 139.8 (d) minutes. The black contours relate to the number density distribution with levels at 7, 25 and 50 $\times10^{9}$ cm$^{-3}$. This clearly shows how shear flow effects induce blob fragmentation and evolution.}
 \label{se}
\end{figure}

The sheared flows that are established by the detailed blob dynamics could also in return influence the further evolution of the condensations. An example is shown in panel a of Fig.~\ref{se}, where we show a signed velocity map, with overlaid contours of the density distribution of condensations at levels of 7, 25 and 50 $\times10^{9}$ cm$^{-3}$ at $t\approx 123.7$ minutes. Concentrating on the density feature labeled with A, after its initial formation, sheared flows already begin to take shape. About 10 minutes later, this segment A is seen as segment A1 and A2 in panel b of Fig.~\ref{se} and the condensation has broken into two distinct segments with increasing separation between them. Segment A2 is also going to break into two segments a little later. At the $t\approx$ 123.7 minutes in panel a of Fig.~\ref{se}, this segment A feature is more like one whole elongated condensation. However, by $t\approx 132.6$ in panel b in Fig.~\ref{se}, several condensations behave totally separate to each other. Another example is the one of segment B in panel a and panel b. This breaks up into segment B1 and segment B2 in panel c at $t\approx$ 136.2 minutes. Then segment B1 further breaks into segment B1 and B3 in panel d at $t\approx$ 139.8 minutes. This gradual change from one elongated dense blob or strand breaking into several segments, surrounded by fast sheared flows, hints at the influence of Kelvin-Helmholtz instabilities (KHI). However, there is no clear vorticity pattern emerging in our simulation that would clearly identify KHI development, which may not have enough time to develop. We speculate that other KHI related substructure may well arise under different parameter settings (field strength, heating scale height), but already establish that sheared flows contribute to the breaking up of elongated condensations into smaller fragments.

\section{Conclusions}
We extended our earlier multidimensional, magnetohydrodynamic simulations of coronal rain occurring in magnetic arcades hosting chromospheric, transition region, and coronal plasma. The main new results can be summarized as follows.

\begin{enumerate}
\item We find that after the initial formation stage of condensations, expansion rebound shock fronts introduced by fast siphon inflows display a fan-shaped structure, typically. The local conditions of where condensations form influences the detailed dynamics and expansion of these rebound shocks, and can lead to asymmetric expansion fronts or only one-sided expansion shock fronts. 
\item We discussed the process of establishing a structured prominence-coronal-transition-region (PCTR) around coronal rain condensations. The strong radiation loss at the boundary of blobs results in local dips in the gas pressure structure at the blob boundary where the temperature sharply rises from 0.01 MK to a coronal temperature of 0.5 MK. 
\item By extending our 2.5D simulation to a longer time of 6 hours, we obtain a secondary cycle of coronal rain in the simulation. This secondary cycle confirms the deductions from previous 1D simulations and observations that by providing consistent and enough energy, coronal rain can form a secondary cycle or even more.  
\item We study the condensation rate in our 2.5D simulation and find the growth of cool mass in the corona to show a good correlation especially with the faster growth rate in the length of condensations in the direction perpendicular to the field lines. This indicates that the growth of cool mass is dominated by the onset of runaway cooling in neighbouring loops. This significantly exceeds the rates obtained in studies of this growth rate in 1D models, purely along field lines, as we also need to understand the expansion speed of onset of runaway cooling in neighbouring loops. By performing detailed quantitative analysis, we also find that no matter what happens in the corona, a constant heating in the chromosphere keeps on evaporating a certain amount of hot plasma into the corona, establishing a mass cycle.  
\item We look into the impact of declining blobs on the transition region, and find that their rebound shocks can spread as upflows from one footpoint to another footpoint. Following the rebound shocks, evaporation driven upflows with a slower velocity refill the loops and heat them to 2.3 MK again. 
\item Plenty of counter-streaming flows are found in our simulation, and we demonstrated several reasons for forming these flows. One is that the extremely low gas pressure area between two neighbouring coronal rain blobs drives strong siphon flows towards it. These shear flows accompany the blobs until they fall into the transition region. 
\item The counter-streaming flows also in return influence the deformation of the blobs, which can break into several segments, starting from an elongated one.

\end{enumerate}

\acknowledgments
The research has been sponsored by an Odysseus grant of
the FWO Vlaanderen. The results were obtained in the KU Leuven GOA project GOA/2015-014 and by the Interuniversity Attraction Poles 
Programme initiated by the Belgian Science Policy Office (IAP P7/08 CHARM). Part
of the simulations used the infrastructure of the VSC - Flemish Supercomputer 
Center, funded by the Hercules Foundation and the Flemish Government - 
Department EWI. We acknowledge fruitful discussions with P. Antolin, and a very helpful referee report.

\normalfont

\end{document}